# Uniform transverse beam profile with a new type nonlinear magnet

Sheng-Dong Gu(谷胜栋) and Wei-Bin Liu(刘渭滨), Key Laboratory of Particle Acceleration Physics and Technology, Institute of High Energy Physics, Chinese Academy of Sciences, 19(B) Yuquan Road, Beijing 100049, China

**Abstract:** In this paper, a new type magnet is proposed and processed to uniform the transverse beam profile. Compeared to the octupole, the new type magnet can prove a similar octupole magnet field in the middle, but the rise rate declines quickly in the edge. So that, a same uniform beam is got with less particles loss. Besides that, a mechanical structure is added to adjust the width of the middle region to satisfy different transverse dimensions, and this whould further reduce particles loss. Some numerical simulations have been done respectively with the octupole and the new type of magnet to show the advantages of the new type magnet.

**PACS**：29.20

## 1 Introduction

The beam with uniform transverse distribution is required in various applications such as irradiation of targets for isotope production, uniform irradiation of detectors for improved efficiency, irradiation of biologicamples and materials. Many methods have been proposed to get an uniform and well-confined beam. One method to get an uniform beam is using nonlinear optics, this method can get an ideal uniform beam, but it requires a combined magnet that is complicated to process. In practice, the octupole is usually be used to replace the combined magnet, but loss of particles in the halo will be caused. Many people have theoretically studied uniformization of the transverse beam profile with multipole magnets [1-4].

Yuri et al. have studied uniformization of the transverse beam profile using nonlinear magnets in detail. They got the relationship of the density distributionons at the locations of the target and the multipole magnet as follows [1]:

$$\rho_t = \rho_0 \left(\frac{dx_t}{dx_0}\right)^{-1} = \frac{\rho_0}{M_{11} - \frac{\alpha_0}{\beta_0} M_{12} - M_{12} \sum_{n=3}^{\infty} \frac{K_{2n}}{(n-2)!} x_0^{n-2}} = \frac{\rho_0}{M_{11} - \frac{\alpha_0}{\beta_0} M_{12} - M_{12} \sum_{n=3}^{\infty} \frac{K_{2n}}{(n-2)!} x_0^{n-2}} \quad (1)$$

Where M is the transmission matrix and $M_{ij}(i,j = 1,2)$ is the element of M, $K_{2n}$ is the 2n-pole integrated strength of the multipole magnet. Where $\rho_0$ and $\rho_t$ are particle density functions at the initial and target, $\beta_0$ and $\beta_t$ are beta functions at the initial and target. According to this equation, the particle distribution can be transformed into a different one at the target by using nonlinear magnets.

Usually, the beam has a Gaussian distribution or similar to Gaussian distribution. Replace the initial density function $\rho_0$ with a Gaussian function and make a Taylor's expansion, the required magnet field is got. All odd-order nonlinear fields are required to transform an ideal Gaussian beam into a uniform beam. Only use an octupole to prove the nonlinesr field, the effect is not so good, and some particles loss will be caused.

## 2 The new type magnet

According to the effect and disadvantage of the octupole, a new type of magent is meticulous designed and processed. The field of an octupole increase rises up along the axes while quadrupole magnet field rise up with an constant speed. If a shelding device is set in the middle of a quadrupole, the rise speed can be changed. With this idea, a new type of magnet is proposed. The field in the middle region of the new type magnet is similar to the octupole magnet field, but the rise rate declines quickly in the edge. With this specialty, the beam can be transformed to uniform with less particle loss. Fig. 1 is the photo and structure of the new type magnet. The magnet consists of four parts: two C-type dipoles, shielding device, and a base. A lead screw is installed on the base to adjust the distance between the two C-type dipoles. The width of octupole-like region can be regulated from 40mm to 80mm with adjusting the distance of the two dipoles from 41mm to 81mm by turning the lead screw. We have proceed a new type magnet and measured the field. Fig. 2 and Fig. 3 show the comparison between measured results and simulation results at different distances between two C-type dipoles. The solid line is the simulation result while the points are measured values. Due to particularity of the magnet, the magnet field is measured with many different points. And the measured points in Fig. 2 and Fig. 3 are integral values of different points along beam direction. The measured results are in agreement with the simulation results at dieffirent distances. The results prove the feasibility of the new type magnet. As discussed, to use the new type magnet for uniformization, it must has a octupole-like field in the middle. A comparison between the fields of an octupole and the new type magnet is shown in Fig. 4. As mentioned, the width of the octupole-like region can be changed according to the demand. The fields of the new type magnet at minimum and maximum distances are ploted in Fig. 4. The new type magnet has a minimum octupole-like width at the minimum distance and vice versa. Comparing with the octupole, the new type magnet almost has the same magnet field in the octupole-like region, but the growth rate of the new type magnet field decrease quickly exceed the middle which is useful to limit particles loss. According to the theory of Yuri et al., beams can be transformed to be uniform with all odd-order nonlinear magnets. It is hard to make a magnet has all the odd-order components, but using a combination of octupole and dodecapole magnet can get a more uniform beam compared with only using the octupole. The field of the new type magnet is also like a combination magnet field of octupole and dipole magnet. The ratio of the octupole strength to the dodecapole strength is not stringent equal to the theory of Yuri et al., but it is helpful.

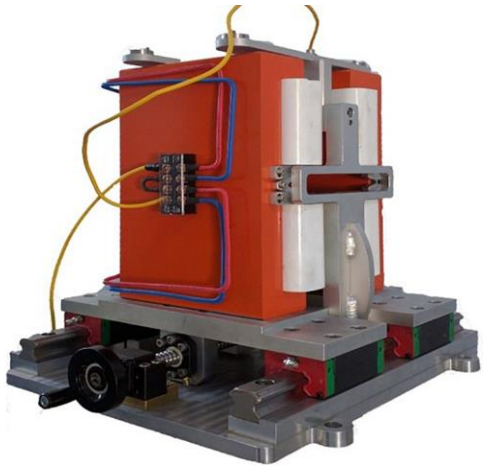

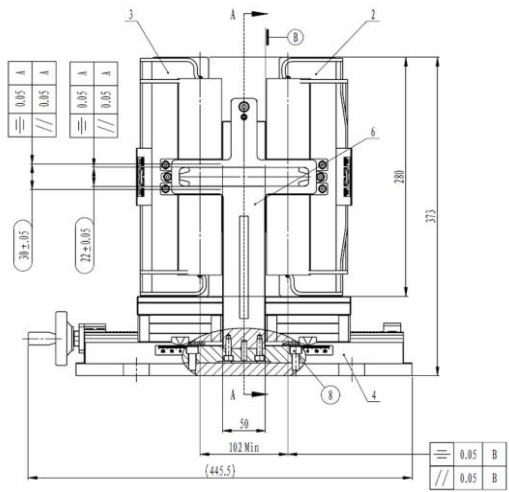

Fig. 1 Photo and structure of the new type magnet

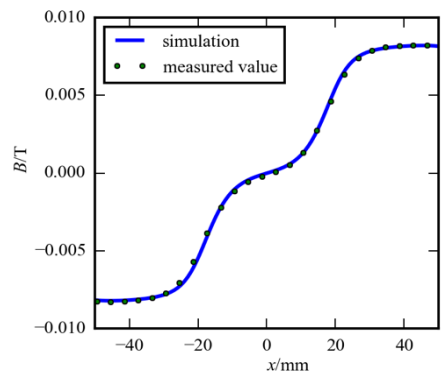

Fig. 2 Comparison between simulation results and measured results while the distance is 41mm

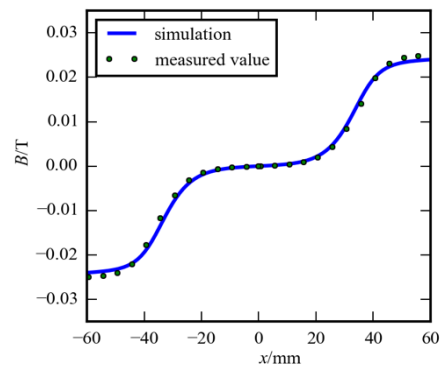

Fig. 3 Comparison between simulation results and measured results while the distance is 61mm

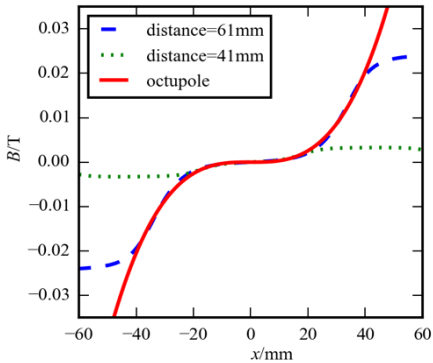

Fig. 4. Comparing between the octupole magnet filed and the new type magnet

## 3 Particle loss comparison between octupole and new type magnet

In this part, particles loss during the uniform process will be discussed with a simple example as shown in Fig. 5. The beam passes a nonlinear magnet and some quadrupoles and get to the target finally. The parameters are listed in Table 1. Nonlinear magnet will result in coupling between horizontal and vertical. In practice, a beam is usually operated to be flat when passing a nonlinear magnet to limit the effect on one direction. In the following, we just discuss one-dimensional case. The basic principle to get a uniform beam is distorting the beam in phase space by changing the momentum with nonlinear magnet. As shown in Fig. 5, when a particle passes an octupole with integral strength KL, the coordinates transformation relations is:

$$\begin{cases} x_1 = x_0 \\ x_1' = x_0' - \dfrac{KL}{6}x_0^3 \end{cases} \quad (2)$$

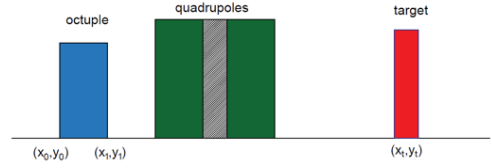

Fig. 5. A schematic plot of the transport system

Where $x_0, x_0'$ and $x_1, x_1'$ are coordinates before and after the octupole. $x'$ is the derivation relative to the position s. The change of the momentum is a cube function of $x_0$.

Table 1 Beam parameters

| Position | α | B[m] | γ[m$^{-1}$] | ε(mm.mrad) | Φ(rad) |
| --- | --- | --- | --- | --- | --- |
| Initial | 53.2 | 28.5 | 40.8 | 3.08 | 0.07 |
| Final | 3.6 | 45.4 | 0.3 | 3.08 | |

Table 2 Hamiltonians and coordinates with differnet magnets at dieffernet locations

| Without nonlinear magnet | | With octupole | | With new type magnet | |
|---|---|---|---|---|---|
| $(x_0, x_0')$ | $H_0$ | $(x_1, x_1')$ | $H_1$ | $(x_1, x_1')$ | $H_1$ |
| [mm, mrad] | [mmmrad] | [mm, mrad] | [mmmrad] | [mm, mrad] | [mmmrad] |
| (10,-14) | 157 | (10,-15) | 251 | (10,-15) | 247 |
| (20,-29) | 630 | (20,-34) | 2647 | (20,-33) | 2287 |
| (30,-43) | 1417 | (30,-60) | 1.6e4 | (30,-54) | 9.8e3 |
| (40,-57) | 2520 | (40,-97) | 6.8e4 | (40,-75) | 2.1e4 |

For a beam, as shown in Fig. 6, the beam is distorted with the effects of dif ferent nonlinear magnets. The two lines plotted in Fig. 6 separately represent distorted beam profiles with octupole and with the new type magnet. The beam size of the profiles in Fig. 6 is up to quadruple of δ. Where δ is root-mean-squared (RMS) radius. The beam profiles are quite different with different magnets. The Hamiltonian can be used to describe the differences. The particle Hamiltonian without and with the octupole can be written as:

$$\begin{cases} H_0 = \gamma x_0^2 + 2\alpha x_0 x_0' + \beta x_0'^2 & (3a) \\ H_1 = \gamma x_0^2 + 2\alpha x_0 \left(x_0' - \frac{KL}{6} x_0^3\right) \\ + \beta \left(x_0' - \frac{KL}{6} x_0^3\right)^2 & (3b) \end{cases}$$

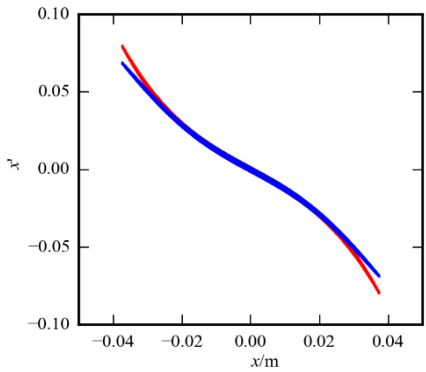

Fig. 6. The nonlinear magnet effect with an octupole (red) and with the new type magnet (blue) (color online)

The Hamiltonian changes at different locations in the beam are listed in Table 2. The changes can be used to describe the beam distortions which is useful to get a uniform beam on the target as discussed above. The Hamiltonian changes with the new type of magnet and the octupole almost the same until to 20mm which is the width of the octupole-like region. 20mm exceed 2δ of this beam and about 95% of particles are located in the range of 2δ for a Gaussian beam. So the new type magnet has the same effect to get a uniform as the octupole. For the particles exceed 2δ, the Hamiltonian changes is huge with an octupole compared with the new type magnet. This means the amplitude of the β oscillation become much bigger for the later transfer, and some particles probably loss because of limit of the pip. The phenomenon can be illustrated with the beam profiles on the target as shown in Fig. 7.

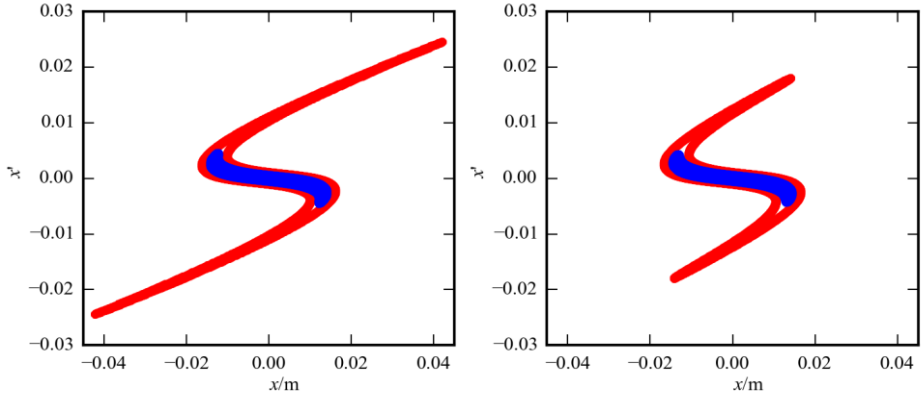

Fig. 7. The transverse beam profile on the target with the octupole (left) and with the new type magnet (right), the green plot the profile with 2δ and the red plot the profile with 4δ (color online)

The left and the right of Fig. 7 show the profiles on the target with an octupole and the new type magnet. With the beam distortion in phase space, the density distribution in the real space is uniform now. Comparing the two pictures, the left one has a long "tail" which is composed by the particles located in quadruple δ. It is because of the overbig Hamiltonian change with an octupole as discussed. This phenomenon is always corresponding with particles in the halo. The long "tail" is usually out of the boundary of the target, and particles in the "tail" become useless. More than that, the particles in the "tail" is difficult to be transported to the target, most of them loss on the pipe during transfer. For the right one, there is no a long "tail" because of the new type magnet has a smaller field from 20mm to 40mm compared to the octupole. In these two figures, for particles in 20mm, the distorted beam profiles have the same shape because of the same changes of Hamiltonian as shown in Table 2. So, the two beams have the same uniform distributions in real space. The advantage of the new type magnet is clear. The same uniform transverse profile beam can be got as that of using the octupole without the "tail", which means no particles in the "tail" will loss in the later transform and almost all of the particles in the beam transformed to the target are in the boundary of target and useful.

## 4  Application

To test the new type magnet, some simulations have been made with the CPHS lattice. Tracewin code is used in these simulations. The parameters of the beam are listed in Table 1, and the β function of the lattice is shown in the Fig. 8. There are two octupoles located in the lattice which is designed for uniformitarian, and a nonlinear magnet is set in one of the locations. 10,0000 macro particles is used in this simulation, including 1% particles in the halo which has tenfold RMS emittance. As discussed, the beam is flat when passes the nonlinear magnet. We just focus on the x direction in these simulations. The octupole strength can be caculated with the theorey of Yuri et al.[1]. The beam density function is Gaussian. According to equation (1), it is easy to calculate the required octupole. In this example, the required integrated strength of the octupole is 5701 $T \cdot m^{-2}$. It should be noted that the strength is calculated with all the odd-order components. Only using the octupole, the octupole stength should be smaller. Comparing the beam distributions on the target, the integrated strength of the octupole is set to 3702 $T \cdot m^{-2}$. Then the new type magnet is adjusted to have the same strength in the middle region compared with the octupole as showing in Fig. 4. The width of the middle region can be got from the size of the transverse profile at the nonlinear magnet position.

The simulation results are compared in many ways. The particle distributions in real space with an octupole and the new type magnet are plotted in Fig. 9 and Fig. 10 to show the uniform effect visually. Both of them have a uniform profile, but there are some particles located far from the core in Fig. 9. Obviously, these particles are part of the "tail" as discussed above and most of them become useless because of the target limit. To compare the uniformity of the two distributions, the number of particles at different locations along x is plotted in Fig. 11.

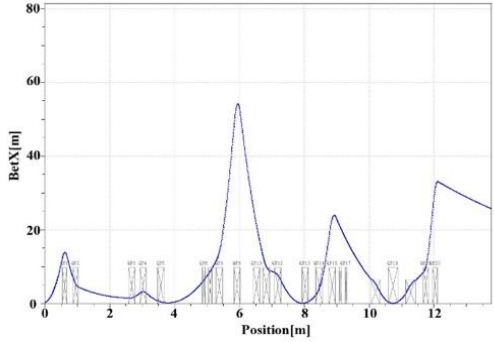

Fig. 8. The $\beta_x$ function of the CPHS lattice

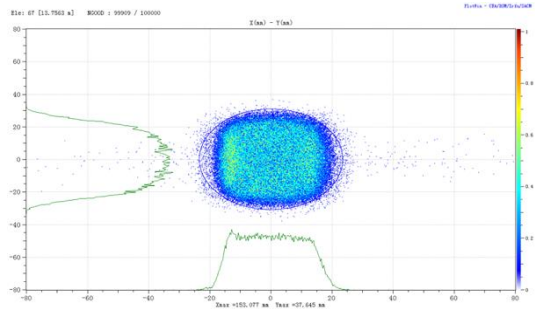

Fig. 9 Particles distribution on the target with an octupole (color online)

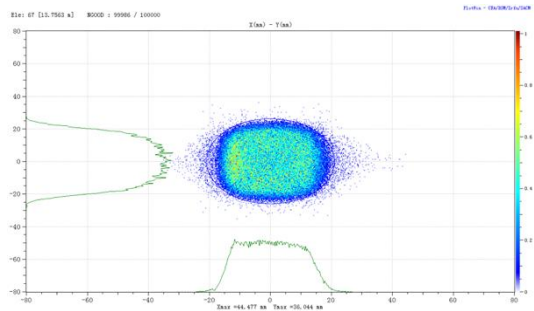

Fig. 10 Particles distribution on the target with the new type mangnet (color online)

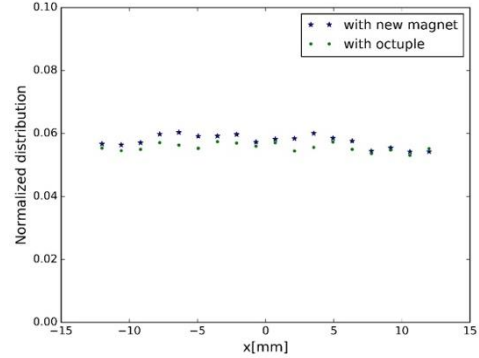

Fig. 11 Particle distribution along x

The variances of these data can be used to present the uniformity of two different distributions. The variances of the two different distributions with the octupole and new type of magnet is 2269 and 1758. Comparing the two variances, the two distributions can be supposed to have the same uniformity. As discussed, some particles in "tail" will get lost. The number of the lost macro particles is listed in Table 3. Comparing these dates, there are 77 more particles loss with an octupole. It is another evidence to prove the advantage of the new type magnet.

Table 3: Comparison of the particles loss

| Type | Macro particles loss |
| --- | --- |
| None | 5 |
| Octupole | 91 |
| The new type magnet | 14 |

## 5 Conclusions

A new type magnet is proposed and designed to replace the octupole with the same uniform effect. We have studied particles loss caused by the octupole while uniforming transverse beam profile and compared with the new type magnet. The new type magnet has two advantages compared with the octupole:

(1) Get the uniform beam with less particle loss.

(2) Adjustable octupole-like region to fit different beam transverse dimensions.